\documentclass[aps,12pt,superscriptaddress,amsfonts,amssymb,amsmath]{revtex4}
\pdfoutput=1

\usepackage{graphicx}
\usepackage{epsfig}
\usepackage{makeidx}
\usepackage{epstopdf}
\usepackage{natbib}
\usepackage{xcolor}
\usepackage{subcaption}
\usepackage{amsmath}

\begin{document}

\title{Quantum uncertainty and energy flux \\ in extended electrodynamics}

\author{F.\ Minotti \footnote{Email address: minotti@df.uba.ar}}
\affiliation{Universidad de Buenos Aires, Facultad de Ciencias Exactas y Naturales, Departamento de F\'{\i}sica, Buenos Aires, Argentina}
\affiliation{CONICET-Universidad de Buenos Aires, Instituto de F\'{\i}sica del Plasma (INFIP), Buenos Aires, Argentina}

\author{G.\ Modanese \footnote{Email address: giovanni.modanese@unibz.it}}
\affiliation{Free University of Bozen-Bolzano \\ Faculty of Science and Technology \\ I-39100 Bolzano, Italy}

\linespread{0.9}

\begin{abstract}

In quantum theory, for a system with macroscopic wavefunction, the charge density and current density are represented by non-commuting operators. It follows that the anomaly $I=\partial_t \rho + \nabla \cdot \mathbf{j}$, being essentially a linear combination of these two operators in the frequency-momentum domain, does not admit eigenstates and has a minimum uncertainty fixed by the Heisenberg relation $\Delta N \Delta \phi \simeq 1$ which involves the occupation number and the phase of the wavefunction. We give an estimate of the minimum uncertainty in the case of a tunnel Josephson junction made of Nb. Due to this violation of the local conservation of charge, for the evaluation of the e.m.\ field generated by the system it is necessary to use the extended Aharonov-Bohm electrodynamics. After recalling its field equations, we compute in general form the energy-momentum tensor and the radiation power flux generated by a localized oscillating source. The physical requirements that the total flux be positive, negative or zero yield some conditions on the dipole moment of the anomaly $I$.

\end{abstract}

\maketitle

\section{Introduction}

The extended electrodynamics theory based on the Aharonov-Bohm lagrangian has attracted much interest over the last years \cite{ohmura1956new, aharonov1963further, van2001generalisation, woodside2009three, jimenez2011cosmological, hively2012toward, Modanese2017MPLB, modanese2017electromagnetic, arbab2017extended, hively2019classical}. Unlike the standard Maxwell theory, the extended electrodynamics allows to compute the  fields generated by physical systems in which the condition of local conservation of charge is not exactly satisfied. Such violations of local conservation are quite rare and may occur especially at a microscopic level; therefore the currents involved are usually small, but the associated physical effects are nevertheless interesting and might lead to useful applications. In our recent work \cite{Minotti-Modanese-Symmetry2021} we computed the  radiation field emitted by oscillating high-frequency currents for which the anomalous  moment $\mathbf{P}$ is not exactly zero, being $\mathbf{P}$ defined as the dipole moment of the ``extra-current'' $I$ that quantifies the anomaly in the local conservation of charge:
\begin{equation}
    I(\mathbf{x},t)=\partial_t \rho(\mathbf{x},t) + \nabla \cdot \mathbf{j}(\mathbf{x},t)
\end{equation}
\begin{equation}
    \mathbf{P}(t)=\int d^3x' \, \mathbf{x}' I(\mathbf{x}',t)
\end{equation}

In order to give an immediate feeling of the formalism involved, let us recall here that the extended field equations with sources are
\begin{equation}
    \nabla \cdot \mathbf{E} =\frac{\rho }{\varepsilon _{0}}-\frac{\partial S}{\partial t}
\end{equation}
\begin{equation}
    \nabla \times \mathbf{B} =\mu _{0}\mathbf{j}+\varepsilon _{0}\mu _{0}\frac{
\partial \mathbf{E}}{\partial t}+\nabla S
\end{equation}
(the equations without sources coincide with the corresponding Maxwell equations). The ``auxiliary'' scalar field $S$ is generated by the extra-current $I$ (eq.\ (\ref{ondaS})) and is zero in the Maxwell theory.

The wave equations for $\mathbf{E}$, $\mathbf{B}$ and $S$ are
\begin{equation}
    \Box \mathbf{E} = -\mu_0 \left( \frac{\partial \mathbf{j}}{\partial t} +c^2 \nabla \rho \right)
    \label{ondaE}
\end{equation}
\begin{equation}
    \Box \mathbf{B} = \mu_0 \nabla \times \mathbf{j}
    \label{ondaB}
\end{equation}
\begin{equation}
    \Box S = \mu_0 I
    \label{ondaS}
\end{equation}
where $\Box$ is the D'Alembert differential operator $(1/c^2)\partial_t^2-\nabla^2$.

In the far-field radiative solutions of eqs.\ (\ref{ondaE}) - (\ref{ondaS}) a longitudinal component of $\mathbf{E}$ is generally present, which of course does not exist in Maxwell theory because $S$ is zero and therefore $\nabla \cdot \mathbf{E}$ is also zero outside the sources. Such anomalous longitudinal component can be expressed in function of $\mathbf{P}$ as $E_L=(\mu_0/4\pi r)\Dot{\mathbf{P}}(t-r/c)\cdot \mathbf{n}$.

In order to assess the physical relevance of the theory, we need to understand under which conditions a violation of local conservation can occur, yielding $I\neq 0$. The main candidates are physical systems of the following types:

\begin{enumerate}
    \item Complex condensed-matter systems described by a quantum field theory, in which the local conservation of the current operator is spoiled by anomalies occurring in the renormalization process \cite{cheng1984gauge,parameswaran2014probing}.

\item Molecular devices, like e.g.\ carbon nanotubes and other molecular ``wires'', in which the effect of bound electrons in the inner orbitals upon the conduction electrons is modelled through a non-local potential, and the anomaly is not due to the use of a reduced eigenstates base, but remains at any order in the computations \cite{walz2015local,jensen2019current,garner2019helical,garner2020three}.

\item Systems with explicitly non-local wave equations, e.g.\ fractional quantum mechanics and other phenomenological models \cite{lenzi2008solutions, lenzi2008fractional, latora1999superdiffusion, caspi2000enhanced, chamon1997nonlocal, balantekin1998green, laskin2002fractional, wei2016comment,modanese2018time}.

\item Macroscopic quantum systems where an uncertainty relation involving the product $\Delta \rho \Delta j$ holds. This is the case that will be considered in this paper.

\end{enumerate}

The idea that quantum uncertainties and quantum tunnelling could spoil the local conservation of charge, which seems classically so unavoidable, was mentioned in some early works on extended electrodynamics \cite{van2001generalisation,hively2012toward}. This intuitive idea is however in conflict with the property of local conservation of probability that is well grounded in the Schr\"odinger equation. In fact, when the number of particles is large and they are incoherent, the real flux of particles follows closely the probability flux; then locally-conserved models of tunnelling and conduction based on the Schr\"odinger equation work well. A typical example is the scanning tunnelling microscope \cite{tersoff1983theory}.

At the other extreme, when the particles number is small and the motion of particles is random and unpredictable, such that the wavefunction only gives a probabilistic description, the interaction of the particles with the e.m.\ field cannot be described through classical field equations, but only considering the probabilities of photon emission etc.

The first issue analysed in this work thus concerns the effect of uncertainties in macroscopic quantum systems like superconductors or superfluids, which can carry currents able to generate a classical e.m.\ field. We shall consider the specific example of a plasma resonance in a Josephson junction and the consequences of the phase-number uncertainty relation $\Delta N \Delta \phi \sim 1$ (Sect.\ \ref{sec2}).

The second main contribution of this work concerns the dynamics of the e.m.\ field in the extended theory, and more precisely its local balance of energy and momentum. For the first time, the density of energy and momentum of the field and their flux are computed in a rigorous and consistent way, through a $T^{ik}$ tensor which respects the usual symmetry requirement. (In Sects.\ \ref{AB-lagr}, \ref{t-tensor} we use Landau-Lifshitz notation with latin indices $i,k...=0,1,2,3$.) As discussed in Sect.\ \ref{relation-with-previously} the expression for the energy density that is obtained directly from the field equations, like in Maxwell theory \cite{van2001generalisation,hively2012toward} gives a mathematically correct relation between the fields $\mathbf{E}$, $\mathbf{B}$, $S$, but does not allow to write consistent expressions for the energy flux and the density of force (generalization of Lorenz force). For this reason we have introduced in Sect.\ \ref{t-tensor} the general definition of the $T^{ik}$ tensor through a coupling with an external gravitational background.

The calculation is quite complex, but the final results for the generalized Lorenz force $\mathbf{f}$ and its power $w$ are remarkably simple (eqs.\ (\ref{eq-w}), (\ref{eq-f})). The new terms in $\mathbf{f}$ and $w$ are respectively equal to $I\mathbf{A}$ and $I\phi$, where $\mathbf{A}$ and $\phi$ are the Aharonov-Bohm potentials. These potentials admit some residual gauge transformations of the form $\mathbf{A} \to \mathbf{A}+\nabla \chi$, $\phi \to \phi+\partial_t \chi$, with $\Box \chi=0$. The conservation laws are invariant with respect to these transformations. Here we have limited ourselves to consider the case of localized oscillating sources for which the potentials are uniquely given by retarded integrals and can be approximately expressed in terms of the standard oscillating dipole moment $\mathbf{p}$ and the anomalous moment $\mathbf{P}$. The total energy flux at infinity can also be explicitly computed and leads to interesting physical 
conditions on the anomalous source (Sects.\ \ref{radiated-power}, \ref{concl}).

Finally we would like to point out that also at the purely classical level the finite-differences technique for numerical solution of the Maxwell equations must deal with the practical impossibility to ensure, in the evaluation of certain matter/field interactions, the exact local conservation of charge \cite{munz2000divergence}.

\section{Quantum uncertainty of local charge conservation in the Josephson plasma resonance}
\label{sec2}

\subsection{Tunnel Josephson junctions and plasma resonance}

We analyse a macroscopic quantum system where the uncertainty relation between the phase of the collective wavefunction and the particle occupation number leads to an uncertainty in the condition of local charge conservation. This system is a tunnel Josephson junction and specifically we consider in the calculation a Nb-NbAlOx-Nb junction made of Niobium and Aluminum oxide, with a critical current $I_J$ of 143 $\mu$A and a capacitance $C$ of 6 pF \cite{gronbech2004microwave}.

In quantum theory this system is described by a wavefunction having a certain amplitude and phase. At the same time, it can be modelled classically as a circuit in which the Josephson junction is a non-linear component, and which also includes a capacitance $C$, an effective inductance $L$ and a resistance $R$ (RCSJ model). The Josephson equations (which in fact have a domain of application much wider than the microscopic BCS theory where they have been originally derived) allow to relate the quantum phase $\phi$ with the supercurrent in the junction. This is essential for our application of the uncertainty relation. An alternative approach, based on the more abstract concept of ``quantum circuit'', was presented in \cite{chen1995quantum,devoret1995quantum}.

For tunnel junctions and other superconducting weak links with capacitance, the Josephson inductance and the capacitance are in parallel. When biased within the supercurrent step ($-I_J<I_0<I_J$), these devices show a damped plasma resonance, in which charge stored on the superconducting surfaces flows backward and forward through the tunnel barrier at frequency $\omega_p=(L_JC)^{-1/2}$, tunable with the bias $I_0$ \cite{waldram1996superconductivity,tinkham2004introduction}. 

The Josephson inductance can be computed as follows: with a DC bias current $I_0<I_J$ there is an equilibrium phase $\phi_0$ determined by the relation $I_0=I_J\sin \phi_0$. 
Consider the Josephson equations
\begin{equation}
    I_s=I_J \sin\phi
    \label{2nd-jos}
\end{equation}
\begin{equation}
   \frac{d\phi}{dt}=\frac{2e}{\hbar}V 
\end{equation}
where $I_s$ is the supercurrent and $\phi$ and $V$ are respectively the phase and voltage differences across the barrier. 

For small deviations from equilibrium we obtain the following relation between  the derivative of the current and the voltage:
\begin{equation}
    \frac{dI_s}{dt}=I_J \cos\phi_0 \frac{d\phi}{dt}=\frac{2eI_J\cos\phi_0}{\hbar}V
\end{equation}
This shows that a small r.f.\ voltage generates a variation in $I_s$, as if the weak link had an effective inductance
\begin{equation}
    L_J=\frac{\hbar}{2eI_J\cos\phi_0}
\end{equation}
which can be tuned by changing $I_0$ and therefore $\phi_0$. The plasma frequency $\omega_p$ is defined as that corresponding to the minimum inductance $L_J^{min}=\hbar/(2eI_J)$.

The complete differential equation of the system in the RCSJ model is
\begin{equation}
    \frac{\hbar C}{2e}\frac{d^2\phi}{dt^2} + \frac{\hbar}{2eR} \frac{d\phi}{dt} = I_0-I_J\sin\phi + I_\Omega \cos(\Omega t)
    \label{eqRCSJ}
\end{equation}
where $I_\Omega$ is the external r.f.\ bias which excites the resonance and $R$ is the normal resistance of the link, that can be considered in parallel to the junction and determines the damping. The values of $C$ and $I_J$ for the junction considered imply $\omega_p \simeq 42$ GHz.

The equation (\ref{eqRCSJ}) is not linear and its solutions are known only in approximate or numerical form; in any case, we are only interested here to know that there is a solution corresponding to the plasma resonance.

\subsection{Quantum description and uncertainty relation}

The microscopic description of the tunnelling process in this kind of junctions was given already by Josephson himself \cite{josephson1962possible}, extending the theory of Cohen et al.\ \cite{cohen1962superconductive}. They assumed that in the context of BCS theory the effect of the barrier may be represented by a small term in the Hamiltonian, called the tunnelling Hamiltonian, of the form
\begin{equation}
    \hat{T}=\sum_{L,R}T_{LR} (c_L^+c_R + c_R^+c_L)
\end{equation}
where the suffixes $L$ and $R$ refer to all the electron states on the left and right sides of the barrier and $T_{LR}$ is a matrix element. It was further assumed that there was superfluid present on both sides of the barrier, with a well-defined phase difference $\phi$. The quantum mechanical treatment then leads to a transition rate proportional to $T_{LR}^2$ and also to $\sin\phi$. 

In general, however, in a superfluid state the phase $\phi$ and the pair number $N$ are conjugate variables, so if we choose a wavefunction whose phase difference is fixed, the allocation of pairs to the two sides of the barrier will be uncertain, and vice versa \cite{waldram1996superconductivity,elion1994direct}. Therefore if we are interested also into the charge density, we need to consider on each side the general uncertainty relation
\begin{equation}
    \Delta \phi \Delta N \simeq 1
    \label{uncert}
\end{equation}

A similar relation holds in quantum optics  between the number of photons in the collective wavefunction and the phase of the wavefunction, at a given position and instant \cite{fox2006quantum}.

In the description of the tunnelling process cited above, $N$ is supposed to be very large. It follows that a large uncertainty $\Delta N$ is acceptable, as long as $\Delta N \ll N$, and the phase $\phi$ can be precisely determined. We shall see, however, that in a Josephson plasma resonance at high frequency the number of oscillating pairs is relatively small and as a consequence the balancing between $\Delta \phi$ and $\Delta N$ is more problematic.

Since $\phi$ has magnitude order 1, we can rewrite (\ref{uncert}) as 
\begin{equation}
    \frac{\Delta \phi}{\phi} \, \frac{\Delta N}{N} \simeq \frac{1}{N}
\end{equation}
At any instant the supercurrent in the junction is connected to the phase by the Josephson equation (\ref{2nd-jos}).
It follows that the uncertainty on the current is $\Delta I_s=I_J\cos\phi \Delta \phi$ and that
\begin{equation}
    \frac{\Delta I_s}{I_s}=\cot \phi \, \frac{\Delta \phi}{\phi}
\end{equation}
During the plasma resonance, the value of $\phi$ is very close to $\phi_0$ defined by the bias current. Therefore except for special values of $\phi_0$ we can simply suppose that $\cot\phi \simeq 1$ as magnitude order, and we obtain
\begin{equation}
    \frac{\Delta I_s}{I_s} \, \frac{\Delta N}{N} \simeq \frac{1}{N}
\end{equation}

\subsection{Charge conservation relation on the electrodes}

Now consider the local conservation relation
\begin{equation}
    \partial_t \rho + \nabla \cdot \mathbf{J}=0
\end{equation}
evaluated on the ``superconducting electrodes''. Since charge oscillates with frequency $\omega_p \sim 10^9$ Hz and the variations in the current density occur (in the tunnelling direction, suppose the $x$-direction) over a length scale $d\sim 10^{-9}$ m, the quantity $\partial_t \rho + \partial_x J_x$ can be approximated, as magnitude order in SI units, as
\begin{equation}
    \partial_t \rho + \partial_x J_x \simeq \pm 10^9 (\rho-J_x)
\end{equation}
where the + sign in front applies if we are at a point and instant where $J_x$ is decreasing, otherwise we have a -- sign.

(The numerical coincidence of $\omega_p$ and $d^{-1}$ assumed above makes the rest of the argument mathematically simpler, but is not necessary, as long as the two quantities are of the same magnitude order; one can introduce an adimensional factor $\chi$ of order 1 and proceed with an expression like $\omega_p(\rho-\chi J_x)$.)

Remember that the total uncertainty in a difference like (\ref{differ}) is given by the sum of the uncertainties of the single terms. Since we know (and it will be confirmed a posteriori) that charge conservation is at least approximately true, we have $\rho\simeq J_x$ and we can write
\begin{equation}
    \frac{\Delta (\rho-J_x)}{\rho} \simeq \frac{\Delta (\rho-J_x)}{J_x} \simeq \frac{\Delta \rho}{\rho} + \frac{\Delta J_x}{J_x}
    \label{differ}
\end{equation}

Consider the relative uncertainties $\frac{\Delta J_x}{J_x}$ and $\frac{\Delta \rho}{\rho}$. They are respectively equal to the relative uncertainties of $I_s$ and $N$:
\begin{equation}
    \frac{\Delta J_x}{J_x}=\frac{\Delta I_s}{I_s}; \qquad  \frac{\Delta \rho}{\rho}=\frac{\Delta N}{N} 
\end{equation}
Taking into account that 
\begin{equation}
    \frac{\Delta J_x}{J_x} \frac{\Delta \rho}{\rho} \, \simeq \frac{1}{N}
\end{equation}
it follows that the total uncertainty (\ref{differ}) is minimum when the two terms are equal. (To prove this, set $u=\Delta J_x/J_x$, $v=\Delta \rho/\rho$, $uv=1/N$; the sum $u+v$ is minimum when $u=v=1/\sqrt{N}$.)

In conclusion, the minimum uncertainty of $(\rho-J_x)$ relative to either $\rho$ or $J_x$ is of order $1/\sqrt{N}$. By re-introducing the factor $\omega_p=d^{-1}=10^9$ the same conclusion holds for the uncertainty of $(\partial_t \rho-\partial_x J_x)$ relative to either $\partial_t \rho$ or $\partial_x J_x$.

Clearly $1/\sqrt{N}$ is in general a small number for a macroscopic system, but for our Josephson junction it is not very small. Suppose that the resonance current is $\simeq I_J$ (but it could even be definitely smaller, for suitable bias, and this reinforces the argument). The charge crossing the junction during a single oscillation is $\simeq 2\pi I_J/\omega_p \simeq 10^{-14}$ C, corresponding to $N\simeq 10^5$ electron pairs. It follows that the relative uncertainty on the local conservation relation is between $10^{-3}$ and $10^{-2}$. 

With an elementary example, suppose $\rho\sim 10^{9}$ C/m$^3$, like in many low-$T_c$ superconductors, $d \sim 10^{-9}$ m, $\omega_p\sim 10^9$ Hz. 
Thus $\partial_t \rho\simeq -\partial_x J_x\simeq 10^{18}$ A/m$^3$, and if we assume for both a relative uncertainty of $10^{-2}$, then their sum will be $(\partial_t \rho+\partial_x J_x) \simeq 10^{18}(1 \pm 0.01 - 1 \pm 0.01) \simeq \pm 10^{16}$ A/m$^3$.

Uncertainties of this kind are completely due to the quantum fluctuations, and are present also if the wavefunction of the system respects the standard continuity condition for the probability flux (as it happens in the BCS theory). In other quantum theories like fractional quantum mechanics or models with non-local potentials, local charge conservation may fail at the level of the probability flux \cite{modanese2018time}.

We are supposing that the source of an e.m.\ field generated by a state with macroscopic wavefunction $\Psi$ is a quantum average on $\Psi$. In particular, for an extra-source $I=\partial_t \rho+\partial_x J_x$ we take the average $\langle \Psi |I| \Psi \rangle $.
The quantity $I$ is essentially (in a frequency-momentum domain) a linear combination of the non-commuting operators $\rho$ and $J_x$; the quantum uncertainty in $I$ originates from those in $\rho$ and $J_x$.
Even if in the quantum theory an operatorial relation $\partial_t \rho=-\partial_xJ_x$ holds, there exist no common eigenstates for the operators $\partial_t \rho$ and $\partial_xJ_x$. Thus quantum noise in $I$ is inevitable and generates fluctuating non-Maxwellian components in the e.m.\ field. 

For the evaluation of field correlations, quantities like $\langle I(x,t)I(x',t') \rangle$ will need to be computed from a microscopic theory.
Note however that in the argument above we did not make any assumption about how exactly the pairs move across the junction, except for supposing that the current is given by Josephson relation, which has been verified with high accuracy in many experiments.

In superconducting systems with intrinsic Josephson junctions and small coherence length, like YBCO, the uncertainty can be larger, because $\sqrt{N}$ is smaller. In that case its estimate becomes more complicated and will be treated in a separate work.

\section{Aharonov-Bohm lagrangian and Extended Electrodynamics (EED) field equations}
\label{AB-lagr}

For later convenience we consider the Aharonov-Bohm lagrangian in a general
four-dimensional space-time, of metric tensor $g_{ik}$. We take a signature
(+,-,-,-) for coordinates ($x^{0},x^{1},x^{2},x^{3}$), which, for the case
of Minkowski metric are, $x^{0}=ct$, and $x^{\alpha }$ the spatial
three-dimensional Cartesian coordinates, with Greek indices taking values $1,2,3
$, and Latin indices values $0,1,2,3$. The (negative) determinant of the
metric tensor is denoted by $g$, and the invariant four-dimensional volume
element $\sqrt{-g}dx^{0}dx^{1}dx^{2}dx^{3}=\sqrt{-g}d\Omega $.

In order to describe the electromagnetic field we take as fundamental
four-vectors for potentials and current (in SI units):%
\begin{eqnarray*}
K_{i} &=&\left( \frac{\phi }{c},-\mathbf{A}\right) , \\
J^{i} &=&\left( \rho c,\mathbf{j}\right) ,
\end{eqnarray*}%
where $\phi $ is the scalar potential, $\mathbf{A}$ the three-dimensional
vector potential, $\rho $ the charge density, and $\mathbf{j}$ the
three-dimensional current vector.

The electromagnetic tensor is%
\begin{equation*}
F_{ik}=\frac{DK_{k}}{Dx^{i}}-\frac{DK_{i}}{Dx^{k}}=\frac{\partial K_{k}}{%
\partial x^{i}}-\frac{\partial K_{i}}{\partial x^{k}},
\end{equation*}%
where $D/Dx^{i}$ represents the covariant derivative, in terms of which the
covariant four-divergence of the four-potential is%
\begin{equation*}
\frac{DK^{m}}{Dx^{m}}=\frac{DK_{m}}{Dx_{m}}=\frac{1}{\sqrt{-g}}\frac{%
\partial }{\partial x^{m}}\left( \sqrt{-g}K^{m}\right) =\frac{1}{\sqrt{-g}}%
\frac{\partial }{\partial x^{m}}\left( \sqrt{-g}K_{l}g^{lm}\right) ,
\end{equation*}%
where%
\begin{equation*}
\frac{D}{Dx_{m}}=g^{lm}\frac{D}{Dx^{l}}.
\end{equation*}

The Aharonov-Bohm lagrangian density is given by 
\begin{equation*}
\Lambda ^{AB}=\Lambda ^{M}+\Lambda ^{\prime },
\end{equation*}%
where%
\begin{equation*}
\Lambda ^{M}=-\frac{1}{4\mu _{0}}F_{ik}F^{ik}=-\frac{1}{4\mu _{0}}%
F_{ik}F_{lm}g^{il}g^{km},
\end{equation*}%
is Maxwell's lagrangian density, and 
\begin{equation*}
\Lambda ^{\prime }=-\frac{1}{2\mu _{0}}\left( \frac{DK^{m}}{Dx^{m}}\right)
^{2}.
\end{equation*}

The Aharonov-Bohm action is thus given by 
\begin{equation*}
S_{AB}=\frac{1}{c}\int \left[ \Lambda ^{M}+\Lambda ^{\prime } -J^{i}K_{i}\right] \sqrt{%
-g}d\Omega ,
\end{equation*}%
whose variation with respect to the four potential $K^{i}$ gives 
\begin{equation*}
\delta S_{AB}=\frac{1}{c}\int \left[ \frac{1}{\mu _{0}}\frac{DF_{ik}}{Dx_{k}}%
-\frac{1}{\mu _{0}}\frac{D}{Dx^{i}}\left( \frac{DK^{m}}{Dx^{m}}\right) +J_{i}%
\right] \delta K^{i}\sqrt{-g}d\Omega ,
\end{equation*}%
and thus%
\begin{equation}
\frac{DF_{ik}}{Dx_{k}}=-\mu _{0}J_{i}+\frac{D}{Dx^{i}}\left( \frac{DK^{m}}{%
Dx^{m}}\right) =-\mu _{0}J_{i}+\frac{DS}{Dx^{i}},  \label{EED_inhom}
\end{equation}%
where we have used the definition of the auxiliary scalar field%
\begin{equation*}
S=\frac{DK^{m}}{Dx^{m}}.
\end{equation*}

Noting that%
\begin{equation*}
\frac{DF_{ik}}{Dx_{k}}=\frac{D^{2}K_{k}}{Dx^{i}Dx_{k}}-\frac{D^{2}K_{i}}{%
Dx^{k}Dx_{k}},
\end{equation*}%
relation (\ref{EED_inhom}) can be alternatively written as 
\begin{equation}
\frac{D^{2}K_{i}}{Dx^{k}Dx_{k}}=\mu _{0}J_{i}.  \label{EED_4pot}
\end{equation}

The so-called homogeneous equations are the same as Maxwell's, resulting
from the definition of the electromagnetic tensor:%
\begin{equation}
\frac{DF_{ik}}{Dx^{l}}+\frac{DF_{li}}{Dx^{k}}+\frac{DF_{kl}}{Dx^{i}}=\frac{%
\partial F_{ik}}{\partial x^{l}}+\frac{\partial F_{li}}{\partial x^{k}}+%
\frac{\partial F_{kl}}{\partial x^{i}}=0  \label{EED_hom}
\end{equation}

In the metric of interest, Minkowski metric, with%
\begin{equation*}
g_{ik}=g^{ik}=\left( 
\begin{array}{cccc}
1 & 0 & 0 & 0 \\ 
0 & -1 & 0 & 0 \\ 
0 & 0 & -1 & 0 \\ 
0 & 0 & 0 & -1%
\end{array}%
\right) ,
\end{equation*}%
if we further consider the three-dimensional electric and magnetic field vectors%
\begin{eqnarray*}
\mathbf{E} &=&-\nabla \phi -\frac{\partial \mathbf{A}}{\partial t}, \\
\mathbf{B} &=&\nabla \times \mathbf{A},
\end{eqnarray*}%
the equations (\ref{EED_inhom}) and (\ref{EED_hom}) reduce in three-dimensional
vector notation to EED equations 
\begin{subequations}
\label{EEDfields}
\begin{eqnarray}
\nabla \cdot \mathbf{E} &=&\frac{\rho }{\varepsilon _{0}}-\frac{\partial S}{
\partial t},  \label{gauss_ext} \\
\nabla \times \mathbf{B} &=&\mu _{0}\mathbf{j}+\varepsilon _{0}\mu _{0}\frac{
\partial \mathbf{E}}{\partial t}+\nabla S,  \label{ampere_ext} \\
\nabla \cdot \mathbf{B} &=&0, \\
\nabla \times \mathbf{E} &=&-\frac{\partial \mathbf{B}}{\partial t},  \label{faraday_ext} 
\end{eqnarray}
\end{subequations}
while from the four-divergence of Eq. (\ref{EED_inhom}) we have
\begin{equation}
\frac{1}{c^{2}}\frac{\partial ^{2}S}{\partial t^{2}}-\nabla ^{2}S=\mu _{0}%
\left[ \frac{\partial \rho }{\partial t}+\nabla \cdot \mathbf{j}\right]
\equiv \mu _{0}I,  \label{D2Seql}
\end{equation}%
in which possible local non-conservation of charge is quantified by the
"extra source" $I$.

The alternative expression (\ref{EED_4pot}) of the in-homogeneous equations
is written in three-dimensional vector notation as
\begin{subequations}
\label{ABpotentials}
\begin{eqnarray}
\frac{1}{c^{2}}\frac{\partial ^{2}\phi }{\partial t^{2}}-\nabla ^{2}\phi 
&=&\frac{\rho }{\varepsilon _{0}} , \\
\frac{1}{c^{2}}\frac{\partial ^{2}\mathbf{A}}{\partial t^{2}}-\nabla ^{2}
\mathbf{A} &=&\mu _{0}\mathbf{j},
\end{eqnarray}
\end{subequations}
which coincide with Maxwell's equations for the potentials in the Lorenz
gauge. The EED equations have thus a residual gauge invariance given by  
\begin{eqnarray*}
\phi  &\rightarrow &\phi -\frac{\partial \chi }{\partial t}, \\
\mathbf{A} &\rightarrow &\mathbf{A}+\nabla \chi ,
\end{eqnarray*}%
for any function $\chi $ satisfying D'Alembert equation%
\begin{equation*}
\frac{1}{c^{2}}\frac{\partial ^{2}\chi }{\partial t^{2}}-\nabla ^{2}\chi =0.
\end{equation*}

\section{Energy and momentum laws derived from the EED field equations}
\label{Energy-Momentum_EED}
In order to determine power emission and interaction of matter and fields in EED we need to derive the energy and momentum conservation laws for this particular theory. These laws have been previously presented \cite{van2001generalisation,hively2012toward,hively2019classical}, and for completeness we also derive them in this section in the usual manner, starting form the field equations. We will show that these laws, although representing correct relations among the fields, are not physically consistent when interpreted as conservation laws. For this reason we derive consistent laws in the following section, directly form the Aharonov-Bohm lagrangian. 

From the scalar product of Faraday's equation, eq. (\ref{faraday_ext}), by $\mathbf{B}/\mu _{0}$, and
of (the extended) Ampere-Maxwell equation, eq. (\ref{ampere_ext}), by $\mathbf{E}/\mu _{0}$ one has%
\begin{eqnarray*}
\frac{\partial }{\partial t}\left( \frac{B^{2}}{2\mu _{0}}\right) &=&-\frac{1%
}{\mu _{0}}\mathbf{B}\cdot \left( \nabla \times \mathbf{E}\right) , \\
\frac{\partial }{\partial t}\left( \frac{\varepsilon _{0}E^{2}}{2}\right) &=&%
\frac{1}{\mu _{0}}\mathbf{E}\cdot \left( \nabla \times \mathbf{B}\right) -%
\mathbf{j}\cdot \mathbf{E}-\frac{1}{\mu _{0}}\mathbf{E}\cdot \nabla S.
\end{eqnarray*}

Adding both equations, and using the identity
\begin{equation*}
\mathbf{E}\cdot \left( \nabla \times \mathbf{B}\right) -\mathbf{B}\cdot
\left( \nabla \times \mathbf{E}\right) =-\nabla \cdot \left( \mathbf{E}%
\times \mathbf{B}\right) ,
\end{equation*}%
together with%
\begin{eqnarray*}
\mathbf{E}\cdot \nabla S &=&\nabla \cdot \left( S\mathbf{E}\right) -S\nabla
\cdot \mathbf{E} \\
&=&\nabla \cdot \left( S\mathbf{E}\right) -\frac{\rho }{\varepsilon _{0}}S+%
\frac{\partial }{\partial t}\left( \frac{S^{2}}{2}\right) ,
\end{eqnarray*}%
where (the extended) Gauss equation, eq. (\ref{gauss_ext}), was used to write the second line, one has a relation that could be considered as an energy conservation law
\begin{equation}
\frac{\partial }{\partial t}\left( \frac{\varepsilon _{0}E^{2}}{2}+\frac{%
B^{2}}{2\mu _{0}}+\frac{S^{2}}{2\mu _{0}}\right) +\nabla \cdot \left( \frac{%
\mathbf{E}\times \mathbf{B}}{\mu _{0}}+\frac{S\mathbf{E}}{\mu _{0}}\right) +%
\mathbf{j}\cdot \mathbf{E}-\frac{\rho S}{\varepsilon _{0}\mu _{0}}=0.
\label{EEED}
\end{equation}

In order to determine a possible expression of the momentum conservation law we start with the usual specific force (per unit volume) on charge-current distributions
\begin{equation*}
\mathbf{f}_{0}=\rho \mathbf{E}+\mathbf{j}\times \mathbf{B},
\end{equation*}%
which using the EED equations (\ref{EEDfields}) can be written in terms of only the fields as
\begin{equation*}
\mathbf{f}_{0}=\varepsilon _{0}\left( \nabla \cdot \mathbf{E}+\frac{\partial
S}{\partial t}\right) \mathbf{E}+\frac{1}{\mu _{0}}\left( \nabla \times 
\mathbf{B}-\varepsilon _{0}\mu _{0}\frac{\partial \mathbf{E}}{\partial t}%
-\nabla S\right) \times \mathbf{B}.
\end{equation*}

Using the relations
\begin{eqnarray*}
\left( \nabla \cdot \mathbf{E}\right) \mathbf{E} &=&\nabla \cdot \left( 
\mathbf{EE}\right) -\left( \mathbf{E}\cdot \nabla \right) \mathbf{E}, \\
\left( \nabla \times \mathbf{B}\right) \times \mathbf{B} &=&\left( \mathbf{B}%
\cdot \nabla \right) \mathbf{B}-\frac{1}{2}\nabla B^{2} \\
&=&\nabla \cdot \left( \mathbf{BB}\right) -\frac{1}{2}\nabla B^{2}, \\
\frac{\partial \mathbf{E}}{\partial t}\times \mathbf{B} &\mathbf{=}&\frac{%
\partial }{\partial t}\left( \mathbf{E}\times \mathbf{B}\right) -\mathbf{E}%
\times \frac{\partial \mathbf{B}}{\partial t} \\
&=&\frac{\partial }{\partial t}\left( \mathbf{E}\times \mathbf{B}\right) +%
\mathbf{E}\times \left( \nabla \times \mathbf{E}\right) \\
&=&\frac{\partial }{\partial t}\left( \mathbf{E}\times \mathbf{B}\right)
-\left( \mathbf{E}\cdot \nabla \right) \mathbf{E}+\frac{1}{2}\nabla E^{2},
\end{eqnarray*}%
we can write ($\mathbf{I}$ is the identity tensor) 
\begin{eqnarray}
\mathbf{f}_{0} &=&\varepsilon _{0}\nabla \cdot \left( \mathbf{EE}-\frac{E^{2}%
}{2}\mathbf{I}\right) +\frac{1}{\mu _{0}}\nabla \cdot \left( \mathbf{BB}-%
\frac{B^{2}}{2}\mathbf{I}\right)  \notag \\
&&-\varepsilon _{0}\frac{\partial }{\partial t}\left( \mathbf{E}\times 
\mathbf{B}\right) +\varepsilon _{0}\frac{\partial S}{\partial t}\mathbf{E}-%
\frac{1}{\mu _{0}}\nabla S\times \mathbf{B}.  \label{f0int}
\end{eqnarray}%
We further use%
\begin{eqnarray*}
\frac{\partial S}{\partial t}\mathbf{E} &\mathbf{=}&\frac{\partial }{%
\partial t}\left( S\mathbf{E}\right) -S\frac{\partial \mathbf{E}}{\partial t}%
, \\
\nabla S\times \mathbf{B} &=&\nabla \times \left( S\mathbf{B}\right)
-S\left( \nabla \times \mathbf{B}\right) ,
\end{eqnarray*}%
so that the last two terms in (\ref{f0int}) can be written as%
\begin{eqnarray*}
&&\varepsilon _{0}\frac{\partial }{\partial t}\left( S\mathbf{E}\right) -%
\frac{1}{\mu _{0}}\nabla \times \left( S\mathbf{B}\right) +\frac{1}{\mu _{0}}%
S\left( \nabla \times \mathbf{B}-\varepsilon _{0}\mu _{0}\frac{\partial 
\mathbf{E}}{\partial t}\right) \\
&=&\varepsilon _{0}\frac{\partial }{\partial t}\left( S\mathbf{E}\right) -%
\frac{1}{\mu _{0}}\nabla \times \left( S\mathbf{B}\right) +\frac{1}{\mu _{0}}%
S\left( \mu _{0}\mathbf{j}+\nabla S\right) \\
&=&\varepsilon _{0}\frac{\partial }{\partial t}\left( S\mathbf{E}\right) -%
\frac{1}{\mu _{0}}\nabla \times \left( S\mathbf{B}\right) +\mathbf{j}S+\frac{%
1}{2\mu _{0}}\nabla S^{2}.
\end{eqnarray*}%
The term $\mathbf{j}S$ suggests to include it in an extended force%
\begin{eqnarray}
\mathbf{f} &=&\rho \mathbf{E}+\mathbf{j}\times \mathbf{B}-\mathbf{j}S  \notag
\\
&=&\nabla \cdot \left[ \varepsilon _{0}\left( \mathbf{EE}-\frac{E^{2}}{2}%
\mathbf{I}\right) +\frac{1}{\mu _{0}}\left( \mathbf{BB}-\frac{B^{2}}{2}%
\mathbf{I}\right) +\frac{S^{2}}{2\mu _{0}}\mathbf{I}\right]  \notag \\
&&-\varepsilon _{0}\frac{\partial }{\partial t}\left( \mathbf{E}\times 
\mathbf{B}-S\mathbf{E}\right) -\frac{1}{\mu _{0}}\nabla \times \left( S%
\mathbf{B}\right) .  \label{finter}
\end{eqnarray}

This expression has some reasonable features, like the Maxwell stress tensor, extended to include a contribution from the scalar. However, an inconsistent feature is the last term, because, by writing it in index notation
\begin{equation*}
\left. \nabla \times \left( S\mathbf{B}\right) \right\vert _{\alpha }=\left.
\nabla \times \left( S\nabla \times \mathbf{A}\right) \right\vert _{\alpha }=%
\frac{\partial }{\partial x^{\beta }}\left[ S\left( \frac{\partial A_{\beta }%
}{\partial x^{\alpha }}-\frac{\partial A_{\alpha }}{\partial x^{\beta }}%
\right) \right] ,
\end{equation*}%
we see that it is the divergence of an antisymmetric tensor, which would thus lead to the non-conservation of angular momentum in a closed system \cite{Landau_fields}.

Another inconsistency is due to the difference in sign of the term $S\mathbf{E}$ inside the time derivative, relative to that in the (extended) Poynting vector in Eq. (\ref{EEED}), which implies that for this component the field energy flow and the field momentum have opposite directions.

As shown in the next section consistent energy and momentum conservation relations can be derived directly from the Aharonov-Bohm lagrangian.

\section{Energy-momentum tensor from the Aharonov-Bohm lagrangian and
conservation laws}
\label{t-tensor}

In order to derive a consistent energy-momentum tensor and energy and
momentum conservation laws we take advantage of the expression of the
Aharonov-Bohm lagrangian in a general four-dimensional metric. This allows
the energy-momentum tensor of the fields, $T_{ik}^{AB}$, to be evaluated as 
\cite{Landau_fields}%
\begin{equation*}
\frac{1}{2}\sqrt{-g}T_{ik}^{AB}=\frac{\partial }{\partial g^{ik}}\left( 
\sqrt{-g}\Lambda ^{AB}\right) -\frac{\partial }{\partial x^{l}}\left[ \frac{%
\partial }{\partial \left( \partial g^{ik}/\partial x^{l}\right) }\left( 
\sqrt{-g}\Lambda ^{AB}\right) \right] .
\end{equation*}

Since $\Lambda ^{M}$ does not depend on $\partial g^{ik}/\partial x^{l}$ the
corresponding tensor is very simply determined using that%
\begin{equation*}
\frac{\partial \sqrt{-g}}{\partial g^{ik}}=-\frac{1}{2}\sqrt{-g}g_{ik},
\end{equation*}%
to obtain the well known result%
\begin{equation*}
T_{ik}^{M}=-\frac{1}{\mu _{0}}\left( F_{il}F_{km}g^{lm}-\frac{1}{4}%
F_{lm}F^{lm}g_{ik}\right) .
\end{equation*}

For the tensor corresponding to $\Lambda ^{\prime }$ we make explicit its
dependence on the metric and its derivatives using that%
\begin{eqnarray*}
\frac{DK^{m}}{Dx^{m}} &=&\frac{1}{\sqrt{-g}}\frac{\partial }{\partial x^{m}}%
\left( \sqrt{-g}K_{i}g^{im}\right)  \\
&=&g^{im}\frac{\partial K_{i}}{\partial x^{m}}+K_{i}g^{im}\frac{\partial \ln 
\sqrt{-g}}{\partial x^{m}}+K_{i}\frac{\partial g^{im}}{\partial x^{m}} \\
&=&g^{im}\frac{\partial K_{i}}{\partial x^{m}}-\frac{1}{2}K_{i}g^{im}g_{kr}%
\frac{\partial g^{kr}}{\partial x^{m}}+K_{i}\frac{\partial g^{im}}{\partial
x^{m}} \\
&=&\frac{1}{2}\left( \frac{\partial K_{i}}{\partial x^{k}}+\frac{\partial
K_{k}}{\partial x^{i}}\right) g^{ik}+\frac{1}{2}\left( K_{i}\delta
_{k}^{r}+K_{k}\delta _{i}^{r}-K_{m}g^{mr}g_{ik}\right) \frac{\partial g^{ik}%
}{\partial x^{r}}.
\end{eqnarray*}%
With this expression, a direct evaluation gives%
\begin{equation*}
T_{ik}^{\prime }=\frac{1}{\mu _{0}}\left[ K_{i}\frac{\partial S}{\partial
x^{k}}+K_{k}\frac{\partial S}{\partial x^{i}}-\left( \frac{S^{2}}{2}+K^{r}%
\frac{\partial S}{\partial x^{r}}\right) g_{ik}\right] .
\end{equation*}

From now on we can specialize the evaluations in the metric of interest, Minkowski metric, and determine the energy and momentum laws by evaluation of the divergence of the energy tensor. 

For the Maxwell tensor we have%
\begin{equation*}
\mu _{0}\frac{\partial T_{ik}^{M}}{\partial x_{k}}=\frac{1}{2}F^{lm}\frac{%
\partial F_{lm}}{\partial x^{i}}-g^{lm}\left( \frac{\partial F_{il}}{%
\partial x_{k}}F_{km}+F_{il}\frac{\partial F_{km}}{\partial x_{k}}\right) ,
\end{equation*}%
and using the homogeneous equations (\ref{EED_hom}) we have%
\begin{equation*}
\mu _{0}\frac{\partial T_{ik}^{M}}{\partial x_{k}}=-\frac{1}{2}F^{lm}\frac{%
\partial F_{il}}{\partial x^{m}}-\frac{1}{2}F^{lm}\frac{\partial F_{mi}}{%
\partial x^{l}}-F^{kl}\frac{\partial F_{il}}{\partial x_{k}}-g^{lm}F_{il}%
\frac{\partial F_{km}}{\partial x_{k}},
\end{equation*}%
of which the first three terms in the rhs clearly cancel out, while Eq. (\ref{EED_inhom}) gives 
\begin{equation}
\frac{\partial F_{km}}{\partial x_{k}}=\mu _{0}J_{m}-\frac{\partial S}{%
\partial x^{m}},  \label{EEDinhomogeneous}
\end{equation}%
so that we finally have%
\begin{equation*}
\frac{\partial T_{ik}^{M}}{\partial x_{k}}=-F_{ik}\left[ J^{k}-\frac{1}{\mu
_{0}}\frac{\partial S}{\partial x_{k}}\right] .
\end{equation*}

For the additional tensor:%
\begin{eqnarray*}
\mu _{0}\frac{\partial T_{ik}^{\prime }}{\partial x_{k}} &=&\frac{\partial
K_{i}}{\partial x_{k}}\frac{\partial S}{\partial x^{k}}+K_{i}\frac{\partial
^{2}S}{\partial x^{k}\partial x_{k}}+\frac{\partial K_{k}}{\partial x_{k}}%
\frac{\partial S}{\partial x^{i}} \\
&&+K_{k}\frac{\partial ^{2}S}{\partial x^{i}\partial x_{k}}-S\frac{\partial S%
}{\partial x^{i}}-\frac{\partial K^{r}}{\partial x^{i}}\frac{\partial S}{%
\partial x^{r}}-K^{r}\frac{\partial ^{2}S}{\partial x^{r}\partial x^{i}} \\
&=&\frac{\partial K_{i}}{\partial x^{k}}\frac{\partial S}{\partial x_{k}}-%
\frac{\partial K_{k}}{\partial x^{i}}\frac{\partial S}{\partial x_{k}}+K_{i}%
\frac{\partial ^{2}S}{\partial x^{k}\partial x_{k}}\text{,}
\end{eqnarray*}%
so that%
\begin{equation*}
\frac{\partial T_{ik}^{\prime }}{\partial x_{k}}=\frac{1}{\mu _{0}}\left(
-F_{ik}\frac{\partial S}{\partial x_{k}}+K_{i}\frac{\partial ^{2}S}{\partial
x^{k}\partial x_{k}}\right) .
\end{equation*}

We thus finally have for the divergence of the complete tensor%
\begin{equation*}
\frac{\partial T_{ik}^{AB}}{\partial x_{k}}=-F_{ik}J^{k}+\frac{1}{\mu _{0}}%
K_{i}\frac{\partial ^{2}S}{\partial x^{k}\partial x_{k}}.
\end{equation*}

Noting that by the taking the four-divergence of (\ref{EED_inhom}) one has%
\begin{equation}
\frac{\partial ^{2}S}{\partial x^{m}\partial x_{m}}=\mu _{0}\frac{\partial
J_{m}}{\partial x_{m}}=\mu _{0}I,  \label{D2SeqI}
\end{equation}%
we end up with%
\begin{equation*}
\frac{\partial T_{ik}^{AB}}{\partial x_{k}}=-F_{ik}J^{k}+K_{i}I.
\end{equation*}

If one considers the fields interacting with matter, the latter described by
an energy-tensor $T_{ik}^{matter}$, energy-momentum conservation requires
that%
\begin{equation*}
\frac{\partial }{\partial x_{k}}\left( T_{ik}^{AB}+T_{ik}^{matter}\right) =0,
\end{equation*}%
and so 
\begin{equation*}
\frac{\partial T_{ik}^{matter}}{\partial x_{k}}g^{im}=-\frac{\partial
T_{ik}^{AB}}{\partial x_{k}}g^{im}=F_{ik}J^{k}g^{im}-K^{m}I
\end{equation*}%
can be considered the local power and force per unit volume on the matter
due to the fields.

In terms of three-dimensional vectors the power of the fields on matter (power lost by the fields) is%
\begin{equation}
w=c\left( F_{0k}J^{k}-K_{0}I\right) =\mathbf{j}\cdot \mathbf{E}-I\phi ,
\label{eq-w}
\end{equation}%
while the force per unit volume on matter is%
\begin{equation}
\mathbf{f}=\rho \mathbf{E}+\mathbf{j}\times \mathbf{B}-I\mathbf{A}.
\label{eq-f}
\end{equation}

An interesting thing to note is that the potentials have a direct effect on
matter when local conservation of charge is not fulfilled.

Having obtained a symmetric tensor, no more problems with conservation of
total angular momentum exist. Besides, the proportionality of (specific)
energy flow and momentum of the fields is automatically satisfied (no more
problems with the difference in signs of the scalar parts found in the
previous section) since%
\begin{eqnarray*}
\frac{\partial T_{0k}^{AB}}{\partial x_{k}} &=&\frac{1}{c}\frac{\partial
T_{00}^{AB}}{\partial t}+\frac{\partial T_{0\alpha }^{AB}}{\partial
x_{\alpha }}, \\
\frac{\partial T_{\alpha k}^{AB}}{\partial x_{k}} &=&\frac{1}{c}\frac{%
\partial T_{\alpha 0}^{AB}}{\partial t}+\frac{\partial T_{\alpha \beta }^{AB}%
}{\partial x_{\beta }},
\end{eqnarray*}%
and in the first relation $T_{0\alpha }^{AB}$ is proportional to the
specific energy flow, while in the second relation $T_{\alpha 0}^{AB}\left(
=T_{0\alpha }^{AB}\right) $ is proportional to the specific momentum.

The explicit expression of the additional tensor $T_{ik}^{\prime }$ in terms
of three-dimensional vectors and scalars is 
\begin{subequations}
\label{Tprimeik}
\begin{eqnarray}
T_{00}^{\prime } &=&\frac{1}{\mu _{0}}\left( \frac{\phi }{c^{2}}\frac{%
\partial S}{\partial t}-\mathbf{A}\cdot \nabla S-\frac{S^{2}}{2}\right) , \\
T_{0\alpha }^{\prime } &=&T_{\alpha 0}^{\prime }=\frac{1}{\mu _{0}c}\left(
\phi \frac{\partial S}{\partial x^{\alpha }}-A_{\alpha }\frac{\partial S}{
\partial t}\right) , \\
T_{\alpha \beta }^{\prime } &=&\frac{1}{\mu _{0}}\left[ -A_{\alpha }\frac{
\partial S}{\partial x^{\beta }}-A_{\beta }\frac{\partial S}{\partial
x^{\alpha }}+\left( \frac{S^{2}}{2}+\frac{\phi }{c^{2}}\frac{\partial S}{
\partial t}+\mathbf{A}\cdot \nabla S\right) \delta _{\alpha \beta }\right] .
\end{eqnarray}
\end{subequations}
The corresponding term in the conservation of the energy relation is 
\begin{equation*}
\frac{\partial T_{0k}^{\prime }}{\partial x_{k}}=\frac{1}{\mu _{0}c}\frac{%
\partial }{\partial t}\left( \frac{\phi }{c^{2}}\frac{\partial S}{\partial t}%
-\mathbf{A}\cdot \nabla S-\frac{S^{2}}{2}\right) -\frac{1}{\mu _{0}c}\nabla
\cdot \left( \phi \nabla S-\mathbf{A}\frac{\partial S}{\partial t}\right) .
\end{equation*}%
The corresponding term for Maxwell's part is the well known expression%
\begin{equation*}
\frac{\partial T_{0k}^{M}}{\partial x_{k}}=\frac{1}{\mu _{0}c}\frac{\partial 
}{\partial t}\left[ \frac{1}{2}\left( \frac{\left\vert \mathbf{E}\right\vert
^{2}}{c^{2}}+\left\vert \mathbf{B}\right\vert ^{2}\right) \right] +\frac{1}{%
\mu _{0}c}\nabla \cdot \left( \mathbf{E}\times \mathbf{B}\right) ,
\end{equation*}%
so that we have the energy density for the fields%
\begin{equation*}
u=\frac{1}{\mu _{0}}\left( \frac{\left\vert \mathbf{E}\right\vert ^{2}}{%
2c^{2}}+\frac{\left\vert \mathbf{B}\right\vert ^{2}}{2}+\frac{\phi }{c^{2}}%
\frac{\partial S}{\partial t}-\mathbf{A}\cdot \nabla S-\frac{S^{2}}{2}%
\right) ,
\end{equation*}%
the energy flow%
\begin{equation}
\mathbf{S}_{u}=\frac{1}{\mu _{0}}\left( \mathbf{E}\times \mathbf{B}-\phi
\nabla S+\mathbf{A}\frac{\partial S}{\partial t}\right) ,  \label{Su}
\end{equation}%
and the energy conservation relation%
\begin{equation}
\frac{\partial u}{\partial t}+\nabla \cdot \mathbf{S}_{u}+\mathbf{j}\cdot 
\mathbf{E}-I\phi =0.  \label{ener_true}
\end{equation}

In order to determine the momentum conservation law we consider the spatial components of the energy-momentum four-divergence%
\begin{equation*}
\frac{\partial T_{\alpha k}^{AB}}{\partial x_{k}}=-F_{\alpha
k}J^{k}+K_{\alpha }I,
\end{equation*}%
which, written in terms of the contravariant components
\begin{equation*}
\frac{\partial T_{\alpha k}^{AB}}{\partial x_{k}}g^{\gamma \alpha
}=-F_{\alpha k}J^{k}g^{\gamma \alpha }+K^{\gamma }I,
\end{equation*}%
can be expanded in terms of three-dimensional magnitudes as (with sum over the $%
\beta $ index)%
\begin{equation*}
-\frac{1}{c}\frac{\partial T_{\alpha 0}^{AB}}{\partial t}+\frac{\partial
T_{\alpha \beta }^{AB}}{\partial x^{\beta }}=-\rho E_{\alpha }-\left( 
\mathbf{j}\times \mathbf{B}\right) _{\alpha }+IA_{\alpha }.
\end{equation*}

Since the Maxwell components are the well known expressions%
\begin{eqnarray*}
T_{\alpha 0}^{M} &=&-\frac{1}{\mu _{0}c}\left( \mathbf{E}\times \mathbf{B}%
\right) _{\alpha }, \\
T_{\alpha \beta }^{M} &=&-\frac{1}{\mu _{0}}\left[ \frac{1}{c^{2}}\left(
E_{\alpha }E_{\beta }-\frac{\left\vert \mathbf{E}\right\vert ^{2}}{2}\delta
_{\alpha \beta }\right) +B_{\alpha }B_{\beta }-\frac{\left\vert \mathbf{B}%
\right\vert ^{2}}{2}\delta _{\alpha \beta }\right] ,
\end{eqnarray*}%
we have, from the relations (\ref{Tprimeik}),%
\begin{eqnarray*}
T_{\alpha 0}^{AB} &=&-\frac{1}{\mu _{0}c}\left( \mathbf{E}\times \mathbf{B}%
-\phi \nabla S+\mathbf{A}\frac{\partial S}{\partial t}\right) _{\alpha }, \\
T_{\alpha \beta }^{AB} &=&-\frac{1}{\mu _{0}}\left[ \frac{1}{c^{2}}\left(
E_{\alpha }E_{\beta }-\frac{\left\vert \mathbf{E}\right\vert ^{2}}{2}\delta
_{\alpha \beta }\right) +B_{\alpha }B_{\beta }-\frac{\left\vert \mathbf{B}%
\right\vert ^{2}}{2}\delta _{\alpha \beta }\right.  \\
&&\left. +A_{\alpha }\frac{\partial S}{\partial x^{\beta }}+A_{\beta }\frac{%
\partial S}{\partial x^{\alpha }}-\left( \frac{S^{2}}{2}+\frac{\phi }{c^{2}}%
\frac{\partial S}{\partial t}+\mathbf{A}\cdot \nabla S\right) \delta
_{\alpha \beta }\right] .
\end{eqnarray*}

In this way, the components of the field momentum density vector $\mathbf{g}$
are%
\begin{equation*}
g_{\alpha }=-\frac{1}{c}T_{\alpha 0}^{AB}=\frac{1}{\mu _{0}c^{2}}\left( 
\mathbf{E}\times \mathbf{B}-\phi \nabla S+\mathbf{A}\frac{\partial S}{%
\partial t}\right) _{\alpha }
\end{equation*}%
so that $\mathbf{g}=\mathbf{S}_{u}/c^{2}$, as it must. The three-dimensional
symmetric tensor $T_{\alpha \beta }^{AB}$ corresponds to the field stress
tensor, let us call it $\sigma _{\alpha \beta }$ ($\overleftrightarrow{%
\sigma }$ in covariant representation), so that the momentum conservation is
written as%
\begin{equation}
\frac{\partial \mathbf{g}}{\partial t}+\nabla \cdot \overleftrightarrow{%
\sigma }+\rho \mathbf{E}+\mathbf{j}\times \mathbf{B}-I\mathbf{A}=0.
\label{momentum_true}
\end{equation}

\section{Relation with the previously derived "conservation laws"}
\label{relation-with-previously}

It is interesting that we have previously derived the "energy conservation
law" (\ref{EEED}) expressed purely in terms of the fields themselves and not
the potentials. To see its relation with the correct law (\ref{ener_true}),
we use (\ref{D2SeqI}) to write%
\begin{eqnarray*}
\mu _{0}I\phi  &=&\phi \left( \frac{1}{c^{2}}\frac{\partial ^{2}S}{\partial
t^{2}}-\nabla ^{2}S\right)  \\
&=&S\left( \frac{1}{c^{2}}\frac{\partial ^{2}\phi }{\partial t^{2}}-\nabla
^{2}\phi \right) -\nabla \cdot \left( \phi \nabla S-S\nabla \phi \right)  \\
&&+\frac{1}{c^{2}}\frac{\partial }{\partial t}\left( \phi \frac{\partial S}{%
\partial t}-S\frac{\partial \phi }{\partial t}\right) .
\end{eqnarray*}%
Using the first of Eqs. (\ref{ABpotentials}), 
\begin{equation*}
\frac{1}{c^{2}}\frac{\partial ^{2}\phi }{\partial t^{2}}-\nabla ^{2}\phi =%
\frac{\rho }{\varepsilon _{0}},
\end{equation*}%
we have%
\begin{equation*}
\mu _{0}I\phi =\frac{\rho S}{\varepsilon _{0}}-\nabla \cdot \left( \phi
\nabla S-S\nabla \phi \right) +\frac{1}{c^{2}}\frac{\partial }{\partial t}%
\left( \phi \frac{\partial S}{\partial t}-S\frac{\partial \phi }{\partial t}%
\right) ,
\end{equation*}%
which, when replaced in (\ref{ener_true}), gives for its lhs%
\begin{eqnarray*}
&&\frac{1}{\mu _{0}}\frac{\partial }{\partial t}\left( \frac{\left\vert 
\mathbf{E}\right\vert ^{2}}{2c^{2}}+\frac{\left\vert \mathbf{B}\right\vert
^{2}}{2}+\frac{S}{c^{2}}\frac{\partial \phi }{\partial t}-\mathbf{A}\cdot
\nabla S-\frac{S^{2}}{2}\right)  \\
&&+\frac{1}{\mu _{0}}\nabla \cdot \left( \mathbf{E}\times \mathbf{B}-S\nabla
\phi +\mathbf{A}\frac{\partial S}{\partial t}\right) +\mathbf{j}\cdot 
\mathbf{E}-\frac{\rho S}{\varepsilon _{0}\mu _{0}},
\end{eqnarray*}%
which, using%
\begin{equation*}
\nabla \cdot \left( \mathbf{A}\frac{\partial S}{\partial t}\right) -\frac{%
\partial }{\partial t}\left( \mathbf{A}\cdot \nabla S\right) =\frac{\partial 
}{\partial t}\left( S\nabla \cdot \mathbf{A}\right) -\nabla \cdot \left( S%
\frac{\partial \mathbf{A}}{\partial t}\right) ,
\end{equation*}%
results in the lhs of (\ref{ener_true}) to be%
\begin{eqnarray*}
&&\frac{1}{\mu _{0}}\frac{\partial }{\partial t}\left( \frac{\left\vert 
\mathbf{E}\right\vert ^{2}}{2c^{2}}+\frac{\left\vert \mathbf{B}\right\vert
^{2}}{2}+\frac{S}{c^{2}}\frac{\partial \phi }{\partial t}+S\nabla \cdot 
\mathbf{A}-\frac{S^{2}}{2}\right)  \\
&&+\frac{1}{\mu _{0}}\nabla \cdot \left( \mathbf{E}\times \mathbf{B}-S\nabla
\phi -S\frac{\partial \mathbf{A}}{\partial t}\right) +\mathbf{j}\cdot 
\mathbf{E}-\frac{\rho S}{\varepsilon _{0}\mu _{0}},
\end{eqnarray*}%
which, since 
\begin{eqnarray*}
\mathbf{E} &=&-\nabla \phi -\frac{\partial \mathbf{A}}{\partial t}, \\
S &=&\frac{1}{c^{2}}\frac{\partial \phi }{\partial t}+\nabla \cdot \mathbf{A},
\end{eqnarray*}
coincides with the lhs of relation (\ref{EEED}). \ Of course, the correct
energy law is (\ref{ener_true}), while (\ref{EEED}), however correct as a
mathematical relation for the fields, does not have the correct
interpretation in terms of energy density, energy flow and power over matter.

Analogously, by writing%
\begin{eqnarray*}
\mu _{0}IA_{\alpha } &=&A_{\alpha }\left( \frac{1}{c^{2}}\frac{\partial ^{2}S%
}{\partial t^{2}}-\nabla ^{2}S\right)  \\
&=&S\left( \frac{1}{c^{2}}\frac{\partial ^{2}A_{\alpha }}{\partial t^{2}}%
-\nabla ^{2}A_{\alpha }\right) -\frac{\partial }{\partial x^{\beta }}\left(
A_{\alpha }\frac{\partial S}{\partial x^{\beta }}-S\frac{\partial A_{\alpha }%
}{\partial x^{\beta }}\right)  \\
&&+\frac{1}{c^{2}}\frac{\partial }{\partial t}\left( A_{\alpha }\frac{%
\partial S}{\partial t}-S\frac{\partial A_{\alpha }}{\partial t}\right) ,
\end{eqnarray*}%
and replacing it in (\ref{momentum_true}), we can reobtain after a direct,
but lengthy evaluation, the relation between fields (\ref{finter}).

\section{Radiated power from a localized source}
\label{radiated-power}

We can now evaluate the power radiated from a localized source in the dipole, long-wave approximation.

The solution of the wave equation for $S$, Eq. (\ref{D2SeqI}), is%
\begin{equation*}
S\left( \mathbf{x},t\right) =\frac{\mu _{0}}{4\pi }\int \frac{I\left( 
\mathbf{x}^{\prime },t^{\prime }\right) }{\left\vert \mathbf{x}-\mathbf{x}%
^{\prime }\right\vert }d^{3}x^{\prime },
\end{equation*}%
with $t^{\prime }=t-\left\vert \mathbf{x}-\mathbf{x}^{\prime }\right\vert /c$%
.

Considering a normal mode $I\left( \mathbf{x}^{\prime },t^{\prime }\right) =%
\widehat{I}\left( \mathbf{x}^{\prime }\right) \exp \left( -i\omega t^{\prime
}\right) $ we can write%
\begin{eqnarray*}
S\left( \mathbf{x},t\right) &=&\frac{\mu _{0}}{4\pi }\int \frac{\widehat{I}%
\left( \mathbf{x}^{\prime }\right) }{\left\vert \mathbf{x}-\mathbf{x}%
^{\prime }\right\vert }\exp \left[ -i\omega \left( t-\left\vert \mathbf{x}-%
\mathbf{x}^{\prime }\right\vert /c\right) \right] d^{3}x^{\prime } \\
&=&\frac{\mu _{0}}{4\pi }\exp \left( -i\omega t\right) \int \frac{\widehat{I%
}\left( \mathbf{x}^{\prime }\right) }{\left\vert \mathbf{x}-\mathbf{x}%
^{\prime }\right\vert }\exp \left( ik\left\vert \mathbf{x}-\mathbf{x}%
^{\prime }\right\vert \right) d^{3}x^{\prime },
\end{eqnarray*}%
where $k=\omega /c$. In this way, with $S\left( \mathbf{x},t\right) =%
\widehat{S}\left( \mathbf{x}\right) \exp \left( -i\omega t\right) $, we have%
\begin{equation*}
\widehat{S}\left( \mathbf{x}\right) =\frac{\mu _{0}}{4\pi }\int \frac{%
\widehat{I}\left( \mathbf{x}^{\prime }\right) }{\left\vert \mathbf{x}-%
\mathbf{x}^{\prime }\right\vert }\exp \left( ik\left\vert \mathbf{x}-\mathbf{%
x}^{\prime }\right\vert \right) d^{3}x^{\prime }.
\end{equation*}

Considering the source $I$ localized about $\mathbf{x}=0$, for a far distant
(relative to the source dimensions) $\mathbf{x}$ position, we have%
\begin{equation*}
\left\vert \mathbf{x}-\mathbf{x}^{\prime }\right\vert \simeq r\left( 1-\frac{%
\mathbf{x}\cdot \mathbf{x}^{\prime }}{r^{2}}\right) =r\left( 1-\frac{\mathbf{%
n}\cdot \mathbf{x}^{\prime }}{r}\right) ,
\end{equation*}%
with $r=\left\vert \mathbf{x}\right\vert $ and where the unit vector in the
direction of the observation point, $\mathbf{n}=\mathbf{x}/r$ was defined .
Also 
\begin{eqnarray*}
\exp \left( ik\left\vert \mathbf{x}-\mathbf{x}^{\prime }\right\vert \right)
&\simeq &\exp \left( ikr\right) \exp \left( -ik\mathbf{n}\cdot \mathbf{x}%
^{\prime }\right) \\
&=&\exp \left( ikr\right) \left( 1-ik\mathbf{n}\cdot \mathbf{x}^{\prime
}\right) ,
\end{eqnarray*}%
where in the second line it was assumed that the wavelength $\lambda =2\pi
/k $ is large compared to the source dimensions. We thus have%
\begin{equation*}
\widehat{S}\left( \mathbf{x}\right) =\frac{\mu _{0}}{4\pi r}\exp \left(
ikr\right) \int \widehat{I}\left( \mathbf{x}^{\prime }\right) \left( 1-ik%
\mathbf{n}\cdot \mathbf{x}^{\prime }\right) d^{3}x^{\prime }.
\end{equation*}

Since even if the charge is not conserved locally, it is conserved globally,
one has that%
\begin{equation*}
\int \widehat{I}\left( \mathbf{x}^{\prime }\right) d^{3}x^{\prime }=0,
\end{equation*}%
so that%
\begin{eqnarray*}
\widehat{S}\left( \mathbf{x}\right) &=&-i\frac{\mu _{0}k}{4\pi r}\exp \left(
ikr\right) \mathbf{n}\cdot \int \widehat{I}\left( \mathbf{x}^{\prime
}\right) \mathbf{x}^{\prime }d^{3}x^{\prime } \\
&\equiv &-i\frac{\mu _{0}k}{4\pi r}\exp \left( ikr\right) \mathbf{n}\cdot 
\widehat{\mathbf{P}},
\end{eqnarray*}%
where the second moment $\widehat{\mathbf{P}}$ of the Fourier amplitude of
the extra-source was defined. We thus have in this approximation,
transforming back to the time domain, 
\begin{eqnarray}
S\left( \mathbf{x},t\right) &=&-\sum\limits_{\omega }i\frac{\mu _{0}\omega 
}{4\pi cr}\exp \left[ i\left( kr-\omega t\right) \right] \widehat{\mathbf{P}}%
\left( \omega \right) \cdot \mathbf{n}  \notag \\
&=&\frac{\mu _{0}}{4\pi cr}\frac{\partial }{\partial t}\sum\limits_{\omega
}\exp \left[ i\left( kr-\omega t\right) \right] \widehat{\mathbf{P}}\left(
\omega \right) \cdot \mathbf{n}  \notag \\
&=&\frac{\mu _{0}}{4\pi cr}\mathbf{\dot{P}}\left( t-r/c\right) \cdot 
\mathbf{n}.  \label{Ssource}
\end{eqnarray}

In the same approximation it is also readily determined (see \cite{Minotti-Modanese-Symmetry2021})
that ($\mathbf{p}$ is the usual electric dipole)%
\begin{eqnarray*}
\phi \left( \mathbf{x},t\right)  &=&\frac{\mu _{0}c}{4\pi r}
\mathbf{\dot{p}}\left( t-r/c\right) \cdot \mathbf{n}, \\
\mathbf{A}\left( \mathbf{x},t\right)  &=&\frac{\mu _{0}}{4\pi r}\left[ 
\mathbf{\dot{p}}\left( t-r/c\right) -\mathbf{P}\left( t-r/c\right) 
\right] , \\
\mathbf{E}\left( \mathbf{x},t\right)  &=&\frac{\mu _{0}}{4\pi r}\left\{ %
\left[ \mathbf{\ddot{p}}\left( t-r/c\right) \times \mathbf{n}\right]
\times \mathbf{n}+\mathbf{\dot{P}}\left( t-r/c\right) \right\} , \\
\mathbf{B}\left( \mathbf{x},t\right)  &=&\frac{\mu _{0}}{4\pi rc}\left[ 
\mathbf{\ddot{p}}\left( t-r/c\right) -\mathbf{\dot{P}}\left(
t-r/c\right) \right] \times \mathbf{n}.
\end{eqnarray*}%
We can thus determine the flux of the extended Poynting vector through a
distant sphere, centered at the dipole, of surface element $d\mathbf{S}%
=r^{2}\sin \theta d\theta d\varphi \mathbf{n}$, so that the instantaneous
emitted power is%
\begin{eqnarray}
W &=&\oint \frac{1}{\mu _{0}}\left( \mathbf{E}\times \mathbf{B}-\phi \nabla
S+\mathbf{A}\frac{\partial S}{\partial t}\right) \cdot d\mathbf{S} \notag \\
&=&\frac{\mu _{0}}{12\pi c}\left[ 2\left\vert \mathbf{\ddot{p}}-
\mathbf{\dot{P}}\right\vert ^{2}+\left( 2\mathbf{\dot{p}}-
\mathbf{P}\right) \cdot \mathbf{\ddot{P}}\right]. \label{power_dipole}
\end{eqnarray}

\section{Conclusions}
\label{concl}

\subsection{Considerations on the gauge freedom of the theory}
An important point of the Aharonov-Bohm (AB) theory is that the potentials are the fundamental fields, which also appear in directly measurable quantities as the power delivered to, and force on matter, eqs. (\ref{eq-w}) and (\ref{eq-f}), respectively. It is thus necessary to address the issue of the theory gauge freedom mentioned at the end of Section \ref{AB-lagr}.

At variance with Maxwell theory, the wave equations for the potentials are uniquely determined in AB theory, eqs. (\ref{ABpotentials}), so that their fundamental solutions in terms of the sources in unbounded space is given by
\begin{subequations}
\label{uniq_pot_sol}
\begin{eqnarray}
\phi \left( \mathbf{x},t\right)  &=&\frac{1}{4\pi \varepsilon _{0}}\int 
\frac{\rho \left( \mathbf{x}^{\prime },t^{\prime }\right) }{\left\vert 
\mathbf{x}-\mathbf{x}^{\prime }\right\vert }d^{3}x^{\prime }, \\
\mathbf{A}\left( \mathbf{x},t\right)  &=&\frac{\mu _{0}}{4\pi }\int \frac{%
\mathbf{j}\left( \mathbf{x}^{\prime },t^{\prime }\right) }{\left\vert 
\mathbf{x}-\mathbf{x}^{\prime }\right\vert }d^{3}x^{\prime },
\end{eqnarray}
\end{subequations}
with $t^{\prime }=t-\left\vert \mathbf{x}-\mathbf{x}^{\prime }\right\vert /c$.

These equations satisfy the conditions that the potentials are zero at all times prior to the turning on of the sources, and at space points where, at the time considered, no information travelling at the speed of light could have arrived from the sources. These "natural" conditions determine that no solution of the wave equation without sources can be added to the potentials given by eqs (\ref{uniq_pot_sol}), because that solution would have to be present before the sources were turned on. On the other hand, the gauge freedom of the theory allows to add sourceless wave solutions to satisfy boundary conditions when it is more practical to work in terms of these conditions than in terms of the actual sources that give rise to the potentials.

The conclusion is that no actual gauge freedom exists in AB theory if the sources are fully known. The limited gauge freedom left is in fact a flexibility of the theory that allows to work in terms of boundary conditions when, from a practical point of view, the actual sources are difficult to determine.

\subsection{Considerations on the possible sources}
From the definition of the dipole moment of the extra source we can obtain a useful relation as
\begin{eqnarray}
\mathbf{P} &=&\int \mathbf{x}I\left( \mathbf{x},t\right) d^{3}x=\int \mathbf{%
x}\left( \frac{\partial \rho }{\partial t}+\nabla \cdot \mathbf{j}\right)
d^{3}x  \notag\\
&=&\frac{d}{dt}\int \mathbf{x}\rho \left( \mathbf{x},t\right) d^{3}x+\int 
\mathbf{x}\nabla \cdot \mathbf{j}d^{3}x  \notag\\
&=&\mathbf{\dot{p}}+\oint \mathbf{x}\left( \mathbf{j}\cdot d\mathbf{S}%
\right) -\int \mathbf{j}\left( \mathbf{x},t\right) d^{3}x.\label{Pexplicit}
\end{eqnarray}

In the extreme case of a dipole with no current, so that, from (\ref{Pexplicit}), $\overset{.}{\mathbf{p}}=\mathbf{P}$, and one has
\begin{equation*}
W=\frac{\mu _{0}}{12\pi c}\mathbf{P}\cdot \mathbf{\ddot{P}}=\frac{\mu
_{0}}{12\pi c}\frac{d}{dt}\left( \mathbf{P}\cdot \mathbf{\dot{P}}
\right) -\frac{\mu _{0}}{12\pi c}\left\vert \mathbf{\dot{P}}
\right\vert ^{2},
\end{equation*}
which for periodic in time, or transient sources has a negative mean value
\begin{equation}
\left\langle W\right\rangle =-\frac{\mu _{0}}{12\pi c}\left\langle
\left\vert \mathbf{\dot{P}}\right\vert ^{2}\right\rangle . \label{power_periodic}
\end{equation} 
In this case 
\begin{equation*}
\int \left\langle \frac{\partial u}{\partial t}\right\rangle d^{3}x=0,
\end{equation*}
so that the matter appears to gain energy from the fields through an incoming energy flux. 

This counter-intuitive phenomenon does not, in principle, involve a non-conservation of energy, because in order to produce either a periodic or a transient dipole without the presence of a current, a non-electromagnetic agent could provide the necessary energy, acting locally on the source. 

In order to further explore this issue we consider the elementary model of a dipole without current, consisting in two point charges of equal, time varying magnitude, but opposite sign, located at fixed positions $a$ and $-a$ on the $z$ axis. The charge density and corresponding extra source are thus given by
\begin{eqnarray*}
\rho \left( \mathbf{x},t\right)  &=&Q\left( t\right) \delta \left( \mathbf{x}%
-a\mathbf{e}_{z}\right) -Q\left( t\right) \delta \left( \mathbf{x}+a\mathbf{e%
}_{z}\right) , \\
I\left( \mathbf{x},t\right)  &=&\dot{Q}\left( t\right) \delta \left( \mathbf{%
x}-a\mathbf{e}_{z}\right) -\dot{Q}\left( t\right) \delta \left( \mathbf{x}+a%
\mathbf{e}_{z}\right) ,
\end{eqnarray*}%
while the potential is
\begin{eqnarray*}
\phi \left( \mathbf{x},t\right)  &=&\frac{1}{4\pi \varepsilon _{0}}\int 
\frac{\rho \left( \mathbf{x}^{\prime },t^{\prime }\right) }{\left\vert 
\mathbf{x}-\mathbf{x}^{\prime }\right\vert }d^{3}x^{\prime } \\
&=&\frac{1}{4\pi \varepsilon _{0}}\left[ \frac{Q\left( t-\left\vert \mathbf{x%
}-a\mathbf{e}_{z}\right\vert /c\right) }{\left\vert \mathbf{x}-a\mathbf{e}%
_{z}\right\vert }-\frac{Q\left( t-\left\vert \mathbf{x}+a\mathbf{e}%
_{z}\right\vert /c\right) }{\left\vert \mathbf{x}+a\mathbf{e}_{z}\right\vert 
}\right] .
\end{eqnarray*}
We thus have
\begin{equation*}
\int I\left( \mathbf{x},t\right) \phi \left( \mathbf{x},t\right) d^{3}x=
\frac{1}{4\pi \varepsilon _{0}}\left[ \frac{2\dot{Q}\left( t\right) Q\left(
t\right) }{\epsilon \rightarrow 0^{+}}-\frac{\dot{Q}\left( t\right) Q\left(
t-2a/c\right) }{a}\right] ,
\end{equation*}
where $\epsilon$ has units of length. Note that the divergent, self-interaction term cancels when time averaged in the case of a transient, or periodic dipole. 

By Taylor developing $Q\left(t-2a/c\right) $:
\begin{equation*}
Q\left( t-2a/c\right) =Q\left( t\right) -\frac{2a}{c}\dot{Q}\left( t\right) +%
\frac{2a^{2}}{c^{2}}\ddot{Q}\left( t\right) -\frac{4a^{3}}{c^{3}}\dddot{Q}%
\left( t\right) +O\left( \frac{a^{4}Q}{c^{5}}\right) ,
\end{equation*}%
we obtain for the time average in transient, or periodic cases
\begin{eqnarray*}
\int \left\langle I\left( \mathbf{x},t\right) \phi \left( \mathbf{x}
,t\right) \right\rangle d^{3}x &=&\frac{1}{4\pi \varepsilon _{0}}\left[ 
\frac{2}{c}\left\langle \dot{Q}^{2}\left( t\right) \right\rangle -\frac{
4a^{2}}{3c^{3}}\left\langle \ddot{Q}^{2}\left( t\right) \right\rangle
+O\left( \frac{a^{4}Q^{2}}{c^{5}}\right) \right]  \\
&=&\frac{1}{4\pi \varepsilon _{0}}\left[ \frac{1}{2a^{2}c}\left\langle
\left\vert \mathbf{P}\right\vert ^{2}\right\rangle -\frac{1}{3c^{3}}
\left\langle \left\vert \mathbf{\dot{P}}\right\vert ^{2}\right\rangle
+O\left( \frac{a^{2}P^{2}}{c^{5}}\right) \right]  \\
&=&\frac{\mu _{0}}{4\pi }\left[ \frac{c}{2a^{2}}\left\langle \left\vert 
\mathbf{P}\right\vert ^{2}\right\rangle -\frac{1}{3c}\left\langle \left\vert 
\mathbf{\dot{P}}\right\vert ^{2}\right\rangle +O\left( \frac{a^{2}P^{2}}{
c^{3}}\right) \right] .
\end{eqnarray*}

In the dipole approximation, $a\rightarrow 0$ with $P$ finite, all terms of order higher than that of the second one go to zero, the second term corresponds to the incoming power, given by eq. (\ref{power_periodic}), while the (divergent in this approximation) first term indicates a large power transferred locally from the source to the fields. This poses a problem, because, although a non-electromagnetic agent can provide the power to the source, the energy conservation relation (\ref{ener_true}) does not include a mechanism that allows the power transferred to the fields to be given back or dissipated, other than that expressed by the term $\mathbf{j}\cdot \mathbf{E}$, which is absent in the model with no current. 

We can thus conclude that the model source considered is not physically possible, even allowing for the presence of non-electromagnetic mechanisms that could set up that source in principle. This does not mean that a similar type of source is excluded. For example, a source of the type considered, but with a ``slow'' increase in the separation $a$, slow in the sense that $\dot{a}/a\ll \left\vert \dot{Q}/Q\right\vert $, so that 
\begin{equation*}
\frac{d}{dt}\int \left\langle u\right\rangle d^{3}x>0,
\end{equation*}
is possible in principle. The increase in energy of the fields and source must of course originate in the non-electromagnetic agent acting on the source.

On the other hand, we can see with a simple example that there is no anomalous behavior when the extra source is due to a current discontinuity without net charge. Since in this case the electric dipole $\mathbf{p}$ is zero, according to eq. (\ref{Pexplicit}) for a closed circuit in which there is a discontinuity in the current $i$ across a gap of width $a$ we have $\left\vert \mathbf{P}\right\vert =ia$. 

In this case the mean radiated power given by the time average of expression (\ref{power_dipole}) is positive and of value
\begin{equation*}
\left\langle W\right\rangle=\frac{\mu _{0}}{4\pi c}\left\langle\left\vert \mathbf{\dot{P}}\right\vert ^{2}\right\rangle=\frac{
\mu _{0}a^{2}}{4\pi c}\left\langle\left( \frac{di}{dt}\right) ^{2}\right\rangle.
\end{equation*}
Note that this is the same expression that would correspond to $3/2$ times the mean power emitted by a normal dipole with conserved current.

\bibliographystyle{unsrt}
\bibliography{mme2}

\begin{thebibliography}{10}

\bibitem{ohmura1956new}
T.~Ohmura.
\newblock A new formulation on the electromagnetic field.
\newblock {\em Progress of Theoretical Physics}, 16(6):684--685, 1956.

\bibitem{aharonov1963further}
Y.~Aharonov and D.~Bohm.
\newblock Further discussion of the role of electromagnetic potentials in the
  quantum theory.
\newblock {\em Physical Review}, 130(4):1625, 1963.

\bibitem{van2001generalisation}
K.J. Van~Vlaenderen and A.~Waser.
\newblock Generalisation of classical electrodynamics to admit a scalar field
  and longitudinal waves.
\newblock {\em Hadronic Journal}, 24(5):609--628, 2001.

\bibitem{woodside2009three}
D.A. Woodside.
\newblock {Three-vector and scalar field identities and uniqueness theorems in
  Euclidean and Minkowski spaces}.
\newblock {\em American Journal of Physics}, 77(5):438--446, 2009.

\bibitem{jimenez2011cosmological}
J.B. Jim{\'e}nez and A.L. Maroto.
\newblock Cosmological magnetic fields from inflation in extended
  electromagnetism.
\newblock {\em Physical Review D}, 83(2):023514, 2011.

\bibitem{hively2012toward}
L.M. Hively and G.C. Giakos.
\newblock Toward a more complete electrodynamic theory.
\newblock {\em International Journal of Signal and Imaging Systems
  Engineering}, 5(1):3--10, 2012.

\bibitem{Modanese2017MPLB}
G.~Modanese.
\newblock {Generalized Maxwell equations and charge conservation censorship}.
\newblock {\em Modern Physics Letters B}, 31:1750052, 2017.

\bibitem{modanese2017electromagnetic}
G.~Modanese.
\newblock Electromagnetic coupling of strongly non-local quantum mechanics.
\newblock {\em Physica B: Condensed Matter}, 524:81--84, 2017.

\bibitem{arbab2017extended}
A.I. Arbab.
\newblock Extended electrodynamics and its consequences.
\newblock {\em Modern Physics Letters B}, 31(09):1750099, 2017.

\bibitem{hively2019classical}
L.M. Hively and A.S. Loebl.
\newblock Classical and extended electrodynamics.
\newblock {\em Physics Essays}, 32(1):112--126, 2019.

\bibitem{Minotti-Modanese-Symmetry2021}
F.~Minotti and G.~Modanese.
\newblock {Are Current Discontinuities in Molecular Devices Experimentally
  Observable?}
\newblock {\em Symmetry}, 13:691, 2021.

\bibitem{cheng1984gauge}
T.-P. Cheng and L.-F. Li.
\newblock {\em Gauge theory of elementary particle physics}.
\newblock Clarendon Press Oxford, 1984.

\bibitem{parameswaran2014probing}
SA~Parameswaran, T~Grover, DA~Abanin, DA~Pesin, and A~Vishwanath.
\newblock Probing the chiral anomaly with nonlocal transport in
  three-dimensional topological semimetals.
\newblock {\em Physical Review X}, 4(3):031035, 2014.

\bibitem{walz2015local}
M.~Walz, A.~Bagrets, and F.~Evers.
\newblock Local current density calculations for molecular films from ab
  initio.
\newblock {\em Journal of Chemical Theory and Computation}, 11(11):5161--5176,
  2015.

\bibitem{jensen2019current}
A.~Jensen, M.H. Garner, and G.C. Solomon.
\newblock When current does not follow bonds: Current density in saturated
  molecules.
\newblock {\em The Journal of Physical Chemistry C}, 123(19):12042--12051,
  2019.

\bibitem{garner2019helical}
M.H. Garner, A.~Jensen, L.O.H. Hyllested, and G.C. Solomon.
\newblock Helical orbitals and circular currents in linear carbon wires.
\newblock {\em Chemical science}, 10(17):4598--4608, 2019.

\bibitem{garner2020three}
M.H. Garner, W.~Bro-J{\o}rgensen, and G.C. Solomon.
\newblock Three distinct torsion profiles of electronic transmission through
  linear carbon wires.
\newblock {\em The Journal of Physical Chemistry C}, 124(35):18968--18982,
  2020.

\bibitem{lenzi2008solutions}
E.K. Lenzi, B.F. de~Oliveira, L.R. da~Silva, and L.R. Evangelista.
\newblock {Solutions for a Schr{\"o}dinger equation with a nonlocal term}.
\newblock {\em Journal of Mathematical Physics}, 49(3):032108, 2008.

\bibitem{lenzi2008fractional}
E.K. Lenzi, B.F. De~Oliveira, N.G.C. Astrath, L.C. Malacarne, R.S. Mendes, M.L.
  Baesso, and L.R. Evangelista.
\newblock Fractional approach, quantum statistics, and non-crystalline solids
  at very low temperatures.
\newblock {\em The European Physical Journal B-Condensed Matter and Complex
  Systems}, 62(2):155--158, 2008.

\bibitem{latora1999superdiffusion}
V.~Latora, A.~Rapisarda, and S.~Ruffo.
\newblock Superdiffusion and out-of-equilibrium chaotic dynamics with many
  degrees of freedoms.
\newblock {\em Physical Review Letters}, 83(11):2104, 1999.

\bibitem{caspi2000enhanced}
A.~Caspi, R.~Granek, and M.~Elbaum.
\newblock Enhanced diffusion in active intracellular transport.
\newblock {\em Physical Review Letters}, 85(26):5655, 2000.

\bibitem{chamon1997nonlocal}
L.C. Chamon, D.~Pereira, M.S. Hussein, M.A.C. Ribeiro, and D.~Galetti.
\newblock Nonlocal description of the nucleus-nucleus interaction.
\newblock {\em Physical Review Letters}, 79(26):5218, 1997.

\bibitem{balantekin1998green}
A.B. Balantekin, J.F. Beacom, et~al.
\newblock Green's function for nonlocal potentials.
\newblock {\em Journal of Physics G: Nuclear and Particle Physics},
  24(11):2087, 1998.

\bibitem{laskin2002fractional}
N.~Laskin.
\newblock {Fractional Schr{\"o}dinger equation}.
\newblock {\em Physical Review E}, 66(5):056108, 2002.

\bibitem{wei2016comment}
Y.~Wei.
\newblock {Comment on ``Fractional quantum mechanics'' and ``Fractional
  Schr{\"o}dinger equation''}.
\newblock {\em Physical Review E}, 93(6):066103, 2016.

\bibitem{modanese2018time}
G.~Modanese.
\newblock Time in quantum mechanics and the local non-conservation of the
  probability current.
\newblock {\em Mathematics}, 6(9):155, 2018.

\bibitem{tersoff1983theory}
J.~Tersoff and D.R. Hamann.
\newblock Theory and application for the scanning tunneling microscope.
\newblock {\em Physical Review Letters}, 50(25):1998, 1983.

\bibitem{munz2000divergence}
C.-D. Munz, P.~Omnes, R.~Schneider, E.~Sonnendr{\"u}cker, and U.~Voss.
\newblock {Divergence correction techniques for Maxwell solvers based on a
  hyperbolic model}.
\newblock {\em Journal of Computational Physics}, 161(2):484--511, 2000.

\bibitem{gronbech2004microwave}
N.~Gr{\o}nbech-Jensen, M.G. Castellano, F.~Chiarello, M.~Cirillo, C.~Cosmelli,
  L.V. Filippenko, R.~Russo, and G.~Torrioli.
\newblock {Microwave-induced thermal escape in Josephson junctions}.
\newblock {\em Physical Review Letters}, 93(10):107002, 2004.

\bibitem{chen1995quantum}
Bin Chen, You~Quan Li, Hui Fang, Zheng~Kuan Jiao, and Qi~Rui Zhang.
\newblock Quantum effects in a mesoscopic circuit.
\newblock {\em Physics Letters A}, 205(1):121--124, 1995.

\bibitem{devoret1995quantum}
M.H. Devoret et~al.
\newblock Quantum fluctuations in electrical circuits.
\newblock {\em Les Houches, Session LXIII}, 7(8):133--135, 1995.

\bibitem{waldram1996superconductivity}
J.R. Waldram.
\newblock {\em Superconductivity of metals and cuprates}.
\newblock IoP, 1996.

\bibitem{tinkham2004introduction}
M.~Tinkham.
\newblock {\em Introduction to superconductivity}.
\newblock Courier Corporation, 2004.

\bibitem{josephson1962possible}
B.D. Josephson.
\newblock Possible new effects in superconductive tunnelling.
\newblock {\em Physics Letters}, 1(7):251--253, 1962.

\bibitem{cohen1962superconductive}
M.H. Cohen, L.M. Falicov, and J.C. Phillips.
\newblock Superconductive tunneling.
\newblock {\em Physical Review Letters}, 8(8):316, 1962.

\bibitem{elion1994direct}
W.J. Elion, M.~Matters, U.~Geigenm{\"u}ller, and J.E. Mooij.
\newblock {Direct demonstration of Heisenberg's uncertainty principle in a
  superconductor}.
\newblock {\em Nature}, 371(6498):594--595, 1994.

\bibitem{fox2006quantum}
M.~Fox.
\newblock {\em Quantum optics: an introduction}, volume~15.
\newblock Oxford University Publishing, Oxford, 2006.

\bibitem{Landau_fields}
L.~D. Landau and E.~M. Lifshitz.
\newblock {\em The classical theory of fields}.
\newblock Pergamon Press, 1971.

\end{thebibliography}

\end{document}